**Magnetic field induced helical mode and topological transitions in a quasi-ballistic topological insulator nanoribbon with circumferentially quantized surface state sub-bands**

Luis A. Jauregui[1,2], Michael T. Pettes[3,⊥], Leonid P. Rokhinson[4,1,2], Li Shi[3,5], Yong P. Chen[4,1,2,*]

[1] Birck Nanotechnology Center, Purdue University, West Lafayette, IN 47907

[2] School of Electrical and Computer Engineering, Purdue University, West Lafayette, IN 47907

[3] Department of Mechanical Engineering, University of Texas at Austin, Austin, TX 78712

[4] Department of Physics and Astronomy, Purdue University, West Lafayette, IN 47907

[5] Materials Science and Engineering Program, University of Texas at Austin, Austin, TX 78712

[⊥] Current address: Department of Mechanical Engineering, University of Connecticut, Storrs, CT 06269

[*] To whom correspondence should be addressed: yongchen@purdue.edu

A topological insulator (TI) nanowire (NW), where the core is insulating and the 2D spin-helical Dirac fermion topological surface states (TSS) are circumferentially quantized into a series of 1D sub-bands, promises novel topological physics and applications. An axial magnetic flux ($\Phi$) through the core drives periodic topological transitions in the surface sub-bands, changing from being all doubly-degenerate with a gapped Dirac point (DP) at integer (including zero) flux quanta ($\Phi_0 = h/e$, with $h$ being the Planck constant and $e$ the electron charge), to contain a topologically-protected, non-degenerate 1D spin helical mode with restored DP at half-integer flux quanta. The resulting magnetoconductance is predicted to exhibit Aharonov-Bohm oscillations (ABO) with maxima occurring alternatively at half-integer or integer flux quanta (referred to as π-ABO or 0-ABO), depending *periodically* on the Fermi wavevector ($k_F$, with period $2\pi/C$, $C$ being the NW circumference). Here, we report a clear observation of such $k_F$-periodic alternations between 0-ABO and π-ABO in $Bi_2Te_3$ TI nanoribbon (NR, a rectangular cross sectional NW) field effect devices, which exhibit quasi-ballistic transport over ~2 μm (as manifested in length-independent conductance, exponential decaying ABO amplitude with increasing temperature (T), and an 1/T dependence of the extracted phase coherence length). The conductances as functions of the gate voltage at half and zero flux quanta also exhibit clear, but anti-correlated oscillations periodic in $k_F$ (with period $2\pi/C$, equivalently when $C$ encloses an integer multiples of Fermi wavelength $2\pi/k_F$), consistent with the circumferentially quantized surface sub-bands. We also extract the minimal Fermi energy and momentum for TSS to emerge out of the bulk valence band, in agreement with the known $Bi_2Te_3$ bandstructure. Nonlocal measurements further confirm the topological origin of the discrete 0/π phase alternation in our observed ABOs. Our demonstration paves the way for future topologically protected devices that may host exotic physics such as Majorana fermions.

**Main**

Three dimensional topological insulators (3D-TIs) are a new class of quantum matter with an insulating bulk and conducting surface states, topologically protected against time-reversal-invariant perturbations (such as scattering by non-magnetic impurities, crystalline defects, and surface roughness)[1,2]. The metallic surface states of TIs have been unambiguously demonstrated by surface sensitive experiments such as angle-resolved photoemission spectroscopy (ARPES)[3,4] and scanning tunneling microscopy (STM)[5]. However, probing the surface states of TI bulk crystals (such as $Bi_2Te_3$, a schematic of its bandstructure is shown in Fig. 1a, where the bulk bands and the topological surface states, TSS, are depicted) by transport measurements has been more challenging because of the non-insulating bulk conduction[6]. Different approaches have been utilized in order to reduce the bulk contribution, such as the fabrication of nanostructured devices based on nanowires/nanoribbons[7-11] and ultrathin films[12,13].

Topological insulator nanowires (TINWs, including the nanoribbons (TINRs) studied here) are topologically analogous to hollow metallic cylinders. The confinement of surface along the circumference ($C$) gives discretely quantized circumferential momentum $k_\perp$ (Fig. 1b) and generates a series of one-dimensional (1-D) surface sub-bands or quantized surface state modes. Unique to TINWs, the spin-helical surface states cause the spin to be tangential to the surface and perpendicularly locked to the momentum, such that a particle (with momentum along the transport axial direction, $k_{//}$, as depicted in Fig. 1b for a TINR) picks up a Berry's phase of $\pi$ due to the $2\pi$ rotation of the spin as it goes around the circumference. When an axial magnetic field (B) is applied (with flux $\Phi = BA$, where A is the cross sectional area), the electron wave function picks up an Aharonov-Bohm (AB) phase of $2\pi\Phi/\Phi_0$ going around the circumference (with flux quanta $\Phi_0 = h/e$, where $h$ is Planck's constant, $e$ is the electron charge). Considering both of these effects (Berry phase and AB phase), the 1-D sub-bands have the following $\Phi$-dependent dispersion (depicted in Fig. 1c):

$$E_l(k_{//}) = \pm\hbar\upsilon_F\sqrt{k_{//}^2 + k_\perp^2}, \text{ with quantized } k_\perp = \Delta k\left(l + 0.5 - \frac{\Phi}{\Phi_0}\right) \qquad (1),$$

where $\hbar$ is the reduced Planck's constant, $\upsilon_F$ is the TSS Fermi velocity, $\Delta k = 2\pi/C$, and $l$ (= 0, ±1, ±2, …) is the angular momentum quantum number.

Two particularly interesting cases occur: 1) when $\Phi$ is even multiples of $\Phi_0/2$ (including $\Phi = 0$), the Dirac point of TSS is gapped (due to the Berry's phase, responsible for the 0.5-shift in the above Eq. 1 for $k_\perp$) and all sub-bands are doubly-degenerate (with two opposite choices of $k_\perp$); and 2) when $\Phi$ is odd multiples of $\Phi_0/2$, the gap recloses due to the AB phase cancelling (modulo $2\pi$) the Berry's phase, and a spin-less (non-degenerate, $k_\perp = 0$) zero-gap 1-D mode[14-19] emerges from the restored Dirac point (DP) (other sub-bands with nonzero $k_\perp$ remain gapped and doubly-degenerate). This 1-D spin helical mode[20] is unique to TINW (absent in, e.g., carbon nanotubes) and predicted to be topologically protected (forbidding 1D backscattering that reverses $k_{//}$) and host Majorana fermions when proximity-coupled to ordinary s-wave superconductors[15,16,21,22]. In both cases, the adjacent sub-bands (quantized surface modes) are separated by an energy gap (measured at $k_{//} = 0$) of $\Delta E = \hbar v_F \Delta k$ (see also schematic Fermi surface depicted in Fig. S1). As depicted in Fig. 1cd, for ballistic TINWs with the Fermi energy ($E_F$) near the Dirac point (up to $|E_F| < \Delta E/2$), the conductance (G) vs. $\Phi$ is dominated by the presence or absence of the helical mode. Therefore, G vs. $\Phi$ (or B) should oscillate with a period of $\Phi_0$ and should reach maxima at odd multiples of $\Phi_0/2$, known as Aharonov-Bohm oscillations with a phase of $\pi$ ($\pi$-ABO)[14,18]. However, if $\Delta E/2 < E_F < \Delta E$, for $\Phi =$ even multiples of $\Phi_0/2$ there is a double-degenerate surface mode, while for $\Phi =$ odd multiples of $\Phi_0/2$ there is only a single helical mode; therefore G vs. $\Phi$ oscillate with a period of $\Phi_0$ and has maxima at even multiples of $\Phi_0/2$, known as Aharonov-Bohm oscillations with zero phase (0-ABO). If $\Delta E < E_F < 3\Delta E/2$, G vs. $\Phi$ should exhibit $\pi$-ABO again and so on. Such alternations between $\pi$-ABO and 0-ABO would occur periodically in $E_F$, with period $\Delta E$ (Fig. 1cd)[14,15]. A schematic of the expected $\Delta G(B)$ (after subtracting an $E_F$-dependent background in G(B)) for different $E_F$'s is depicted in Fig. 1d. Previously, ABO with a period of $\Phi_0$ and G(B) maxima at even multiples of $\Phi_0/2$ (0-ABO) has been measured in TINRs[7,8,10], and the phase (0) of these ABO were insensitive to gate voltage[8] ($V_g$). Theories have suggested[14] disorder, which broadens the surface sub-bands and makes the number of modes for even or odd multiples of $\Phi_0/2$ unidentifiable, as the reason for ABO phase being not dependent on $V_g$.

Here, we demonstrate gate-tunable ABO with $k_F$-periodic phase alternation between 0 and $\pi$ in quasi-ballistic $Bi_2Te_3$ NR field effect devices. We synthesize single crystal $Bi_2Te_3$ nanoribbons (NRs) by catalyst-free vapor solid method as described previously[23]. NRs are grown out of the plane of the substrate and are individually transferred using an electrochemically sharpened tungsten probe and under an optical microscope to 500 μm thick $SrTiO_3$ (STO) substrates for further studies. STO has been previously utilized as a dielectric substrate to gate TI thin films and nanoribbons [13,23,24]. All data presented below were measured in a representative device with NR width = 150 nm and thickness = 60 nm as measured by atomic force microscopy (AFM). Fig. 2a shows a scanning electron microscope (SEM) image of this device with a schematic of its cross section shown in the inset. Qualitative similar data have been measured in several other devices.

*Field effect*

Effective gate control of TSS is highly desired for investigations of novel quantum transport and device applications of TIs. Figure 2b shows the ambipolar field effect in the 4-probe conductance (G) measured with current I = 1 nA at T = 0.25 K for different segments of the TINR depicted in Fig. 2a. Here $G_{i-j,\ k-l}$ denotes the 4-probe conductance measured with current leads (i, j) and voltage probes (k, l). By varying $V_g$, carriers are tuned from p-type to n-type, with a minimum of G for $V_g$ ~ -2.5V ($V_{CNP}$, the charge neutrality point). The measured G vs. $V_g$ for two segments of the TINR, with different channel lengths ($L_{ch}$), $L_{ch}$ = 1.6 μm and 2.8 μm for $G_{6-1,\ 3-2}$ and $G_{6-1,\ 4-3}$ respectively, are similar in magnitude, suggesting a quasi-ballistic transport. It is also notable that for $V_g$ ~ $V_{CNP}$, G is ~ $e^2/h$ for both segments. The finite G measured likely originate from the invasive electrodes on the TINR[25].

*Aharonov-Bohm oscillations*

Figure 2c depicts the low-temperature (T = 0.25K) magnetoconductance (ΔG(B) = G(B) – f(B), where f(B) is a smooth background subtracted from the raw data G(B)) vs. B-field (depicted in Fig. 2a as the orange arrow) parallel to the NR axis, at three representative $V_g$'s. ΔG(B) oscillates periodically with B, with a period ΔB = 0.48T for all $V_g$'s. The measured ΔB agrees with the expected period for ABO of $\Phi_0/A$ = 0.48T (where A ~ 9000 nm$^2$ is the measured TINR cross sectional area). At $V_g$ = -5V < $V_{CNP}$ ($E_F < E_F^0$, inside the bulk valence band, where $E_F^0$ is the energy of the top of the bulk valence band), we observe ΔG(B) maxima at even multiples of $\Phi_0/2$ (integer flux quanta) and minima at odd multiple of $\Phi_0/2$, as previously observed (0-ABO)[7,8]. These 0-ABO observed here might originate from the topologically-trivial surface states (2D electron gas) formed due to bulk band bending[26] for $V_g < V_{CNP}$, noting the bulk valence band is expected to dominate the conduction for $E_F < E_F^0$ as depicted in the bandstructure of $Bi_2Te_3$ (Fig. 1a). As we increase $V_g$ thus raise $E_F$ into TSS, another type of ABO becomes observable. For example, for $V_g$ ~ $V_{CNP}$ ~ -2.5V ($E_F$ ~ $E_F^0$), we observe ABO with ΔG(B) *maxima* at *odd* multiples of $\Phi_0/2$ (*half*-integer flux quanta) and minima at even multiples of $\Phi_0/2$ (except at Φ = 0), in agreement with the predicted π-ABO. The small peak at B = 0T can be attributed to weak anti-localization[9,23,27-29] (WAL), which likely also occurs and contributes to the peak at B = 0T observed for other $V_g$'s. At $V_g$ = -0.5V, we observe ΔG peaks primarily at even multiples of $\Phi_0/2$ but also a few peaks (especially at $\Phi_0/2$) at odd multiples of $\Phi_0/2$, suggesting a competition between 0-ABO (more dominant at this $V_g$) and π-ABO.

Figure 2d shows a color plot of ΔG vs. parallel B-field and $V_g$ at T = 0.25K. For -2.5V < $V_g$ < 0V we observe that ΔG peaks at alternative values of even or odd integers of $\Phi_0/2$. Such an alternation is particularly notable between the two more prominent peaks at Φ = 0 and Φ = $\Phi_0/2$. However, for $V_g$ < -2.5V (below CNP) the ΔG peak at Φ = $\Phi_0/2$ vanishes, while the peak at Φ = 0 gets very prominent. The

ΔG peak at $\Phi = \Phi_0/2$ is the first oscillation from the π-ABO and is the strongest since it may be less affected by the Zeeman energy compared to the higher odd multiples of $\Phi_0/2$ (that can be more easily dephased at higher B-field in spin-orbit coupled systems[30-32]). Also, this ΔG peak at $\Phi = \Phi_0/2$ (unique to TSS[20]) vanishing for $V_g < V_{CNP}$ ~ -2.5V is consistent with the $Bi_2Te_3$ bandstructure, where the DP of TSS is buried inside the BVB and the BVB would dominate the conduction below the charge neutrality point (CNP, close to the top of BVB).

Figure 3a depicts a higher resolution color map of ΔG vs. parallel B-field and $V_g$ (B-Field step = 4mT, $V_g$ step = 100mV) for $-7V < V_g < +10V$ (see also Fig. S2 for a zoomed-in view over a smaller range of B and V), where the gate induced alternation between 0-ABO and π-ABO is again observed (for $V_g$ down to $V_0$,~ -1.9V slight above the $V_{CNP}$ in this measurement). Figure 3b shows ΔG vs. $V_g$ (vertical cuts from Figure 3a) for the two special $\Phi = 0$ and $\Phi_0/2$, plotted in a relative small $V_g$ range for clarity. Both ΔG ($V_g$) curves exhibit clear oscillations, but the two sets of oscillations are ~180° out of phase, where the maxima (minima) in one curve tend to occur at the minima (maxima) of the other curve. We label each ΔG peak [dip] in the data measured at $\Phi = \Phi_0/2$ [$\Phi = 0$] with an integer oscillation index (N, increasing for increasing order of $V_g$), starting with N = 0 (first ΔG peak [dip]) at $V_g = V_0$ (below which neither the out-of-phase oscillations between the two ΔG ($V_g$) curves nor the alternation between π-ABO and 0-ABO are evident), while we label each ΔG dip [peak] as N + 1/2 (see examples in Figure 3b). For $V_g > V_0$ (note $V_0$ is associated with the $V_g$ required to reach the top of the BVB, labeled in Fig. 1a as $E_F^0$, above which the surface state conduction dominates), we can relate $V_g$ with $k_F$ of TSS using:

$$n_s = C_{ox}(V_g - V_0)/e = ((k_F)^2 - (k_F^0)^2)/4\pi \qquad (2),$$

where $n_s$ is the surface carrier density, $C_{ox}$ is the STO capacitance, and $k_F^0$ is the $k_F$ at the top of the BVB (also note that with relatively large number of transverse $k_\perp$ modes occupied in our samples for $V_g > V_0$, the standard results for 2D spin-helical fermions can be used to relate the gate-induced TSS carrier density with $k_F$, see more discussions in SI Fig. S1). The observed $V_g$ vs. N can be well fitted (solid curves in Figure 3c) to Eq. 2 (assuming $k_F = k_F^0 + N\Delta k$, where $\Delta k = 2\pi/C = 0.0015$Å$^{-1}$ represents the quantized momentum encircling the TINR with $C = 420$ nm the TINR circumference), from which we obtain $C_{ox}$ ~ 100 nF/cm$^2$ and $k_F^0$ ~ 0.06 Å$^{-1}$. We have also checked that introducing an uncertainty of as large as +/- 0.5V in the choice of N = 0 peak do not significantly change extracted $C_{ox}$ and $k_F^0$. The extracted $C_{ox}$ is within the range of the STO capacitance measured at low-temperatures in nanodevices[13,23,33]. We plot the extracted $k_F = (4\pi C_{ox}/e(V_g - V_0) + (k_F^0)^2)^{1/2}$ vs. N in the inset of Figure 3c to demonstrate the oscillation (indexed by N) is periodic in $k_F$. The $k_F^0$ ~ 0.06 Å$^{-1}$ agrees with the minimum

TSS momentum for the energy to exceed the top of bulk valence band (BVB, depicted in Fig. 1a) measured by ARPES[3,4] on bulk $Bi_2Te_3$. For $k_F < k_F^0$, BVB states would dominate the TSS in conduction, and neither the π-ABO nor $\Delta k$-periodic $\Delta G$ ($V_g$) oscillations due to TSS sub-bands can be observed. Also, assuming the linear $E_F$ vs. $k_F$ dispersion, we extract $E_F^0 = \hbar v_F k_F^0 \sim 133$meV, in excellent agreement with the ARPES-measured energy separation between BVB top and Dirac point, called $E_3$ in ref. 4. Therefore, the observed π-ABO and the gate-tunable phase (0 or π) in our TINRs are strong transport experimental evidences of the TSS and the formation of the 1D helical mode in TINRs.

*Ballistic transport from the temperature (T) dependence of AB oscillations*

The T-dependence of the ABO was known to exhibit different behaviors in the diffusive and ballistic regimes from previous studies in metallic and semiconducting rings[34]. The amplitude of ABO in diffusive TINRs showed a ~ $T^{-1/2}$ dependence[7,8], however for ballistic $Bi_2Se_3$ TINRs, the amplitude was found to decay exponentially[10] with T. Fig 4a shows the color plot of $\Delta G$ vs. parallel B-field and T for 0.25K < T < 5K at $V_g$ = -0.78 V. We observe the ABO amplitude decreases and the ABO phase changes with increasing temperature. Figure 4b shows horizontal cuts ($\Delta G$ vs. B) of Figure 4a at a few representative T's, where the change of phase is clearly observed (from π-ABO to 0-ABO). For example, at T = 0.25K, π-ABO is dominant, however when T is raised to T ~ 3.5K the oscillations become 0-ABO. On the other hand, for $V_g$ = -1.2V (Figure S3), 0-ABO are dominant at T = 0.3K, while the oscillations contain more π-ABO ($\Delta G$ peaks mainly at odd multiples of $\Phi_0/2$) at T ~ 2.3K. This change of phase with T may be related to a similar mechanism as dephasing by disorder[14,18]. In order to analyze the T-dependence of ABOs, we compute the fast Fourier transform (FFT, an example of which for T = 0.25K is depicted in the inset of Fig. 4c) of the $\Delta G$ vs. B at different T's. From the FFT we observe clear peaks at frequencies corresponding to periods h/e, h/2e and h/3e in magnet flux. The $n^{th}$ harmonic of the fundamental period (h/e) is due to electrons that travel around the circumference of the TINR n times[34]. The probability of the electron to enclose the TINR circumference multiple times decreases with n, therefore the FFT amplitudes drops with increasing n. The T-dependent amplitudes of FFT peaks (A(FFT)) for the 3 corresponding periods (h/e, h/2e and h/3e) are depicted in Fig. 4c. We observe these amplitudes (obtained by integrating the h/ne FFT peaks in intervals indicated by red horizontal lines in the inset of Fig. 4c) decrease exponentially with increasing temperature ( $A(FFT) \sim e^{-b_n T}$ ) as previously reported for rings made from ballistic two-dimensional electron gases (2DEG)[34] and TINRs[10]. While theories have predicted the decay rates $b_n = nC/(2TL_\Phi)$, where $L_\Phi$ is the phase coherence length, previous experiments in quasi-ballistic devices has shown that $b_n$ is not always linearly dependent on n and this deviation was attributed to thermal averaging[10,34]. In our devices, we have found both types of behaviors, depending on

$V_g$ (e.g. $b_n$ is not proportional to $n$ here at $V_g$ = -0.78V, but $b_n$ is proportional to $n$ at $V_g$ = -1.2V, shown in Fig. S3). Hence, we calculate $L_\Phi$ from each $n^{th}$ harmonic as $L_\Phi = nC/(2Tb_n)$, shown in Fig. 4d (where the average $L_\Phi$ is also plotted). The extracted $L_\Phi$'s for each $n^{th}$ harmonic of $h/e$ (and their average) shows a similar T-dependence ($L_\Phi \sim T^{-1}$), with average $L_\Phi$ reaching as large as 3μm at T ~ 0.25K (this is also consistent with Fig. S3, showing $L_\Phi$ for $V_g$ = -1.2V). It is expected that in the ballistic regime[34], where fermions are weakly coupled to the environment, decoherence is dominated by fluctuations of the environment, resulting in $L_\Phi \sim T^{-1}$. In contrast, previous measurements on more disordered, diffusive TI NWs or NRs[7,27-29] and 2DEG rings[35-37] have all found $L_\Phi \sim T^{-\alpha}$ with α ~ 0.4 - 0.5. Obtaining such a large $L_\Phi$ ~ 3.0 μm (comparable or larger than channel length) at low T and a distinctive exponential T-dependence of the amplitude of ABOs (in contrast to previously studied more disordered and diffusive TINRS[7,8,23], where $A(FFT) \sim T^{-1/2}$) are strong signatures of quasi-ballistic transport in our TINRs. Our measured $L_\Phi$ ~ 3.0 μm at low-T is also consistent with the fact that we can still observe ABO from $\Delta G_{6\text{-}1, 4\text{-}3}$ with $L_{ch}$ ~ 3μm (Figure S4, where due to the longer channel length $L_{ch}$, the ABO was more easily dephased with increasing B-field).

*Non-local Aharonov-Bohm oscillations*

Previous experiments in 2DEG rings showed that the ABO phase could be shifted with $V_g$ when only one arm of the ring was gated (such asymmetric gating induces a potential difference between two arms)[38]. While the Onsager principle could cause only discrete phases (0 or π) to be seen in "local" electrical measurements (as in our experiments, $G_{6\text{-}1, 3\text{-}2}$ or $G_{6\text{-}1, 4\text{-}3}$) under this scenario, when non-local electrical measurements were performed, a continuous change of ABO phase was observed[38-40]. We also performed non-local electrical measurements in our TINRs (device depicted in Fig. 2a). Figure 5a displays the non-local resistance $R_{2\text{-}1, 3\text{-}4}$ and $R_{3\text{-}4, 2\text{-}1}$ vs. $V_g$ at T = 0.25 K, where an ambipolar field effect is also observed, with significant values (~3 kΩ) of non-local resistances at CNP ($V_g$ ~ -2V). The two non-local $R_{2\text{-}1, 3\text{-}4}$ and $R_{3\text{-}4, 2\text{-}1}$ are similar in magnitude, where the separation between the current and voltage probes $L_{ch} = L_{2\text{-}3}$ = 1.6 μm. However, the non-local resistance $R_{4\text{-}5, 3\text{-}2}$, with $L_{ch} = L_{3\text{-}4}$ = 2.8 μm, is considerably reduced compared to $R_{2\text{-}1, 3\text{-}4}$. Such a large reduction (by more than 10 times) of non-local resistance when $L_{ch}$ is increased only by factor ~ 2 may be related to quasi-ballistic transport. Figure 5b displays the color plot of non-local $\Delta G_{2\text{-}1, 3\text{-}4}$ vs. parallel B-field and $V_g$ at T = 0.25K (horizontal color-coded dashed lines correspond to $\Delta G_{2\text{-}1, 3\text{-}4}$ vs. parallel B-field at representative $V_g$'s shown in Fig S5). The non-local ABO period of oscillations $\Delta B$ ~ 0.48T is consistent with the TINR cross sectional area and the period observed in the ABO from local measurements (Fig. 2). The amplitude of ABO ($\Delta G$) measured in the non-local configuration (~ 1 $e^2/h$), about one order of magnitude larger, than that of the ABO measured in the local configuration (~ 0.1 $e^2/h$), is the largest ABO amplitude reported in TIs so far. Similar enhancement of the ABO amplitude was previously measured in 2DEG rings using non-local configuration[38]. Most importantly, ΔG(B) measured in our TINRs using the non-local configuration with different $V_g$'s only gives discrete ABO phases (0 or π only) as in the local configuration, contrary to the continuous change of ABO phase previously observed in asymmetrically gated 2DEG rings. Therefore, we can rule out asymmetric gating as responsible for the discrete change of ABO phase (between 0 or π) tuned by $V_g$ in our TINRs, observed in both local and non-local configurations.

*Conclusions*

We have demonstrated quasi-ballistic transport in TINRs from the channel length ($L_{ch}$)-insensitive conductance and field effect, and the exponentially decaying T-dependence of the ABO amplitude. In both local and nonlocal magnetoconductance measurements, we observed π-ABOs and 0-ABOs alternating periodically with $k_F$ (tuned by $V_g$, with period $\Delta k = 2\pi / C$, C being NR circumference). The conductance oscillations in $V_g$ (periodic in $k_F$ and observed even at B = 0) reveal quantized surface state modes, from which we extracted the minimal $k_F$ for the Fermi energy to cross only TSS and found good agreement with that measured by ARPES. These observations are unambiguous experimental demonstrations of the periodic modulation of surface sub-bands by a magnetic flux Φ threaded parallel to the TINR axis, particularly the existence of the topologically protected 1-D helical mode when Φ = odd multiple of $\Phi_0/2$. Such a helical mode in TINRs may be important for developing topologically protected electronic or spintronic devices and for hosting Majorana fermions.

**Acknowledgements**

The TI material synthesis, characterization and magneto-transport studies are supported by DARPA MESO program (Grant N66001-11-1-4107). Part of the FET fabrication and characterizations are supported by Intel Corporation. L.A.J. acknowledges support by an Intel PhD fellowship and the Purdue Center for Topological Materials fellowship. L.P.R. acknowledges support by DOE BES DE-SC0008630.

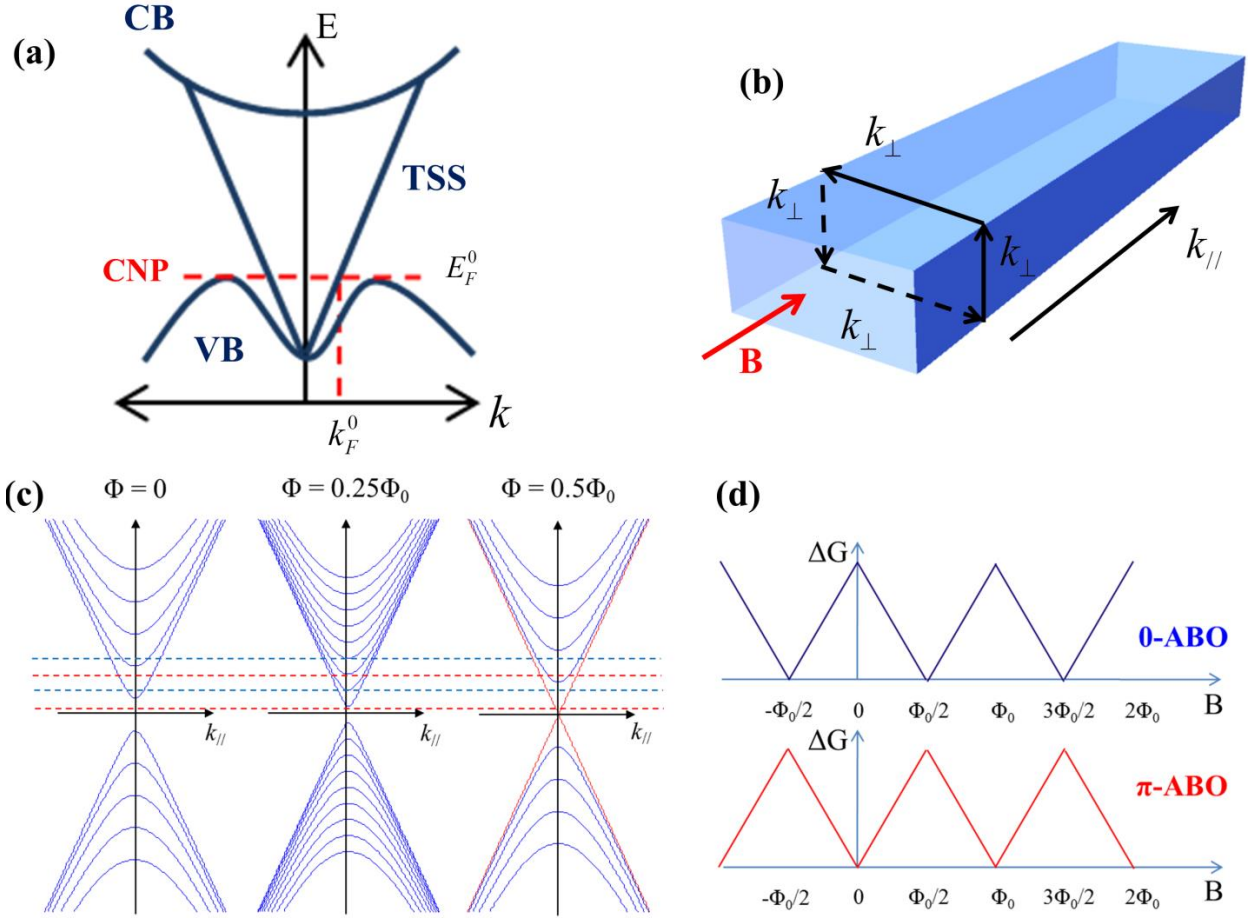

Figure 1. **Schematics of the $Bi_2Te_3$ bandstructure, the surface state modes and expected magnetoconductance oscillation patterns in topological insulator nanoribbons.** (a) Schematic band diagram of bulk Bi2Te3. The bulk conduction band (CB), topological surface states (TSS) and bulk valence band (VB) are labeled. Note the Dirac point (DP) is buried inside the VB, thus only n-type TSS are accessible inside the bulk bandgap. The horizontal and vertical dashed lines mark the minimum Fermi energy ($E_F^0$, at top of VB) and momentum ($k_F^0$) to observe surface state conduction (conduction by the bulk valence band states would dominate for lower $E_F$ or $k_F$). The charge neutrality point (CNP) also occurs close to $E_F^0$. (b) Schematic of a topological insulator nanoribbon (TINR), where $k_{//}$ and $k_\perp$ label TSS momentum parallel and perpendicular to the TINR axis. The applied axial magnetic field (B) is depicted by a red arrow. (c) Schematic of the (circumferentially quantized) TINR surface state modes or surface sub-bands (neglecting bulk bands) depicted for 3 representative axial magnetic fluxes (Φ, in unit of magnetic flux quanta $\Phi_0 = h/e$) through the NR cross section. For each Φ, the multiple surface state modes (sub-bands) arise from discrete quantized $k_\perp$ (see Eq. 1). (d) Schematic of the expected

magnetoconductance (ΔG(B)) oscillation pattern at two different types of $E_F$ positions (exemplified by the blue and red horizontal dashed lines in **c**) giving rise to 0-Aharonov-Bohm oscillations (0-ABO) and π-ABO, respectively.

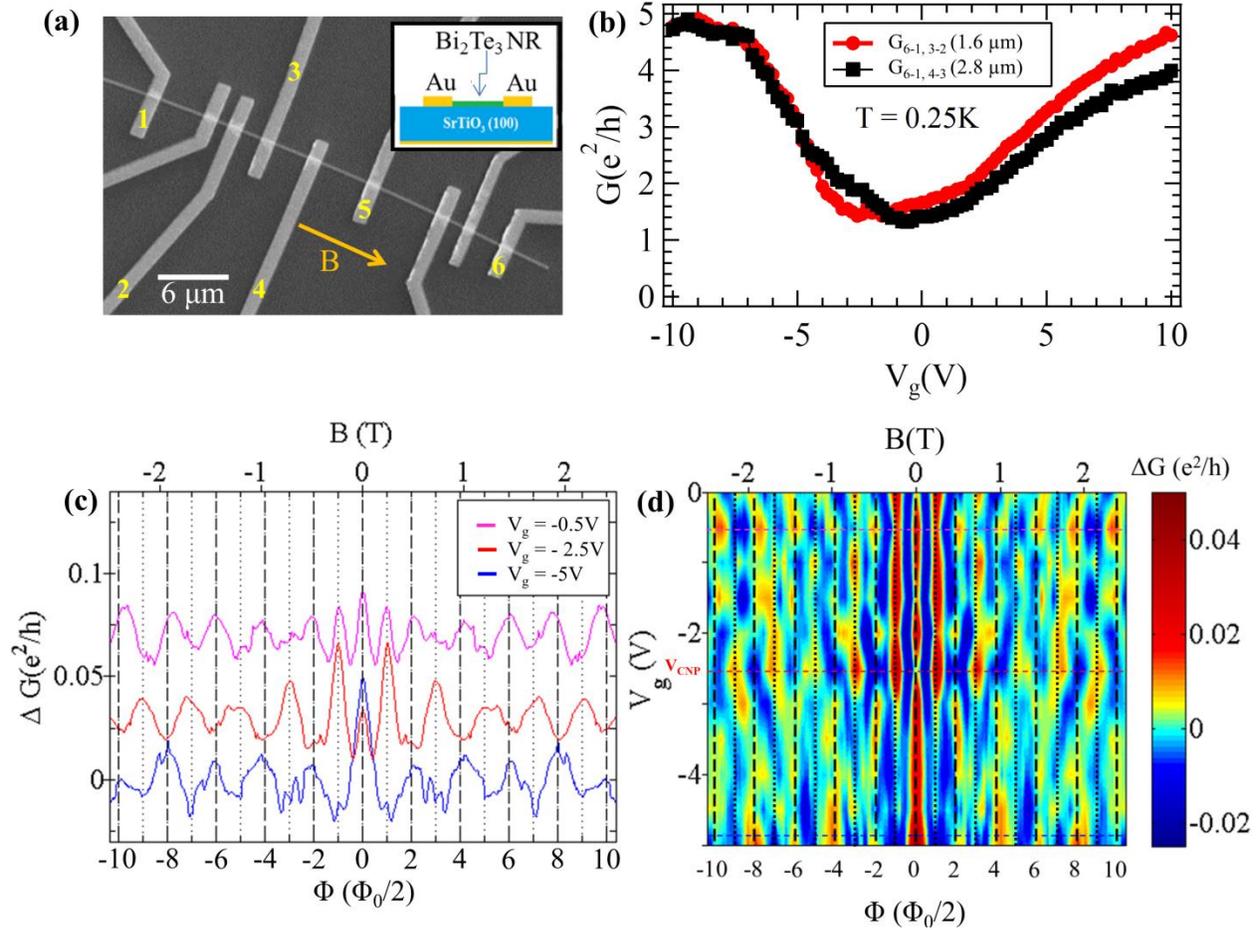

Figure 2. **Ambipolar field effect, demonstrating quasi-ballistic conduction, and gate-tunable 0- and π- ABOs, demonstrating topological surface state modes in TINRs.** (a) Scanning electron microscope (SEM) image of a 150 nm wide, 60 nm thick TINR multi-terminal device on SrTiO$_3$ (STO) substrate studied in this work. The electrodes are numerically labeled. The orange arrow indicates the direction of the B-field applied, along the NR axis. Inset: schematic of the device cross-section. (b) Four-terminal conductance (G) vs. back gate voltage ($V_g$), measured for two different segments of the TINR (between voltage probes 3-2, with length 1.6 μm, and between 4-3, with length 2.8 μm, respectively). The current I = 1 nA is applied between electrodes 6 and 1 (ground). (c) Magnetoconductance (ΔG(B), with a smooth background subtracted), in units of $e^2/h$, vs. B-field (top axis, with corresponding magnetic flux Φ in units of half-flux-quantum ($\Phi_0/2 = h/2e$) in bottom axis) at different $V_g$'s. Curves are vertically offset for clarity.

**(d)** Color plot of ΔG (in units of $e^2/h$) vs. $V_g$ (in 0.5V step) and B (top axis, corresponding Φ in bottom axis). The horizontal color-coded dashed lines correspond to the same color-coded ΔG in Fig. 2c. Vertical dashed/dotted lines in (**c,d**) mark integer/half-integer flux quanta (even/odd multiples of $\Phi_0/2$). Data in (**b-d**) were all measured at temperature T = 0.25K. The conductance in (**c,d**) as well as in Figs. 3 & 4 below were measured between electrodes 3-2. Magnetoconductance data in Figs. 2 and 3 have been symmetrized between opposite B field directions.

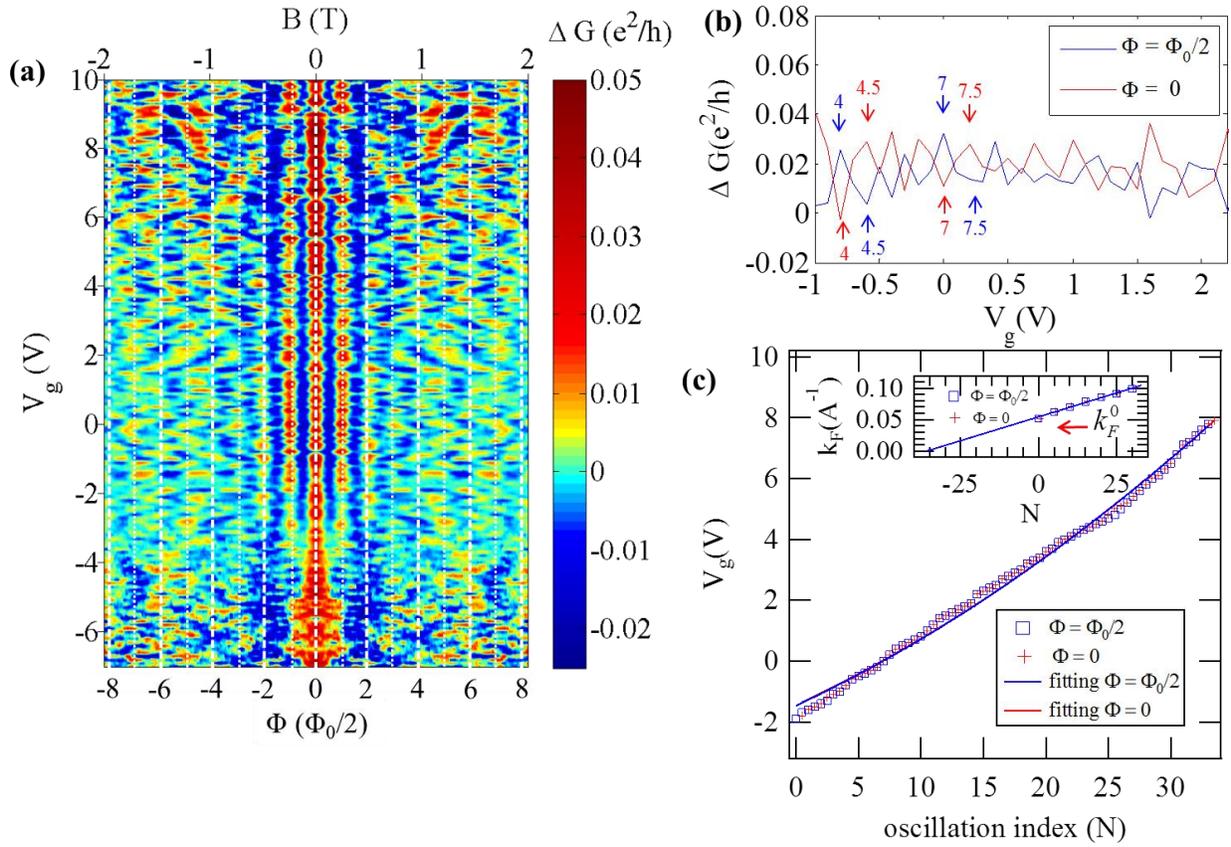

Figure 3. **Analysis of quantized topological surface state sub-bands in gate-dependent conductance oscillations and 0/π-ABO alternations.** (**a**) Color plot of ΔG (in units of $e^2/h$) vs. $V_g$ (in 0.1V step) and B (top axis, corresponding magnetic flux Φ in bottom axis). The vertical dashed lines represent integer flux quanta (Φ = even multiples of $\Phi_0/2$), while vertical dotted lines represent half-integer flux quanta (Φ = odd multiples of $\Phi_0/2$). A zoomed-in view (for $V_g$ from -1 to 2V and for Φ from $-\Phi_0$ to $\Phi_0$) is shown in Fig. S2. (**b**) ΔG (in units of $e^2/h$) vs. $V_g$ (only showing from -1V to 2.2 V for clarity) for Φ = 0 and $\Phi_0/2$, taken from corresponding vertical cuts in Figure 3a. Peaks/dips [dips/peaks] in ΔG ($V_g$) curve at Φ = $\Phi_0/2$ [Φ = 0] are assigned (in increasing order of $V_g$ starting from $V_0$) with consecutive integer/half-integer oscillation indices (N, some examples labeled with arrows). We assign N = 0 to the first peak [dip]

observed at $V_g = V_0 = -1.9V$ (corresponding to the top of valence band) in $\Delta G$ for $\Phi = \Phi_0/2$ [$\Phi = 0$]. **(c)** $V_g$ of the $\Delta G$ peaks/dips vs. N for $\Phi = 0$ and $\Phi = \Phi_0/2$. The solid curves are fittings of $V_g$ vs. N using equation 2. Inset: extracted Fermi momentum ($k_F$) vs. N for $\Phi = 0$ and $\Phi = \Phi_0/2$ (only select data points with N in increment of 5 are shown for clarity, starting from N = 0; solid line is a linear fitting). The extracted $k_F$ for N = 0 ($k_F^0$) ~ 0.06 Å$^{-1}$ is in reasonable agreement with the minimal momentum (depicted in Fig. 1a) for TSS to appear above the top of BVB measured by ARPES[4] in $Bi_2Te_3$. T = 0.25 K for all measurements in (a-c).

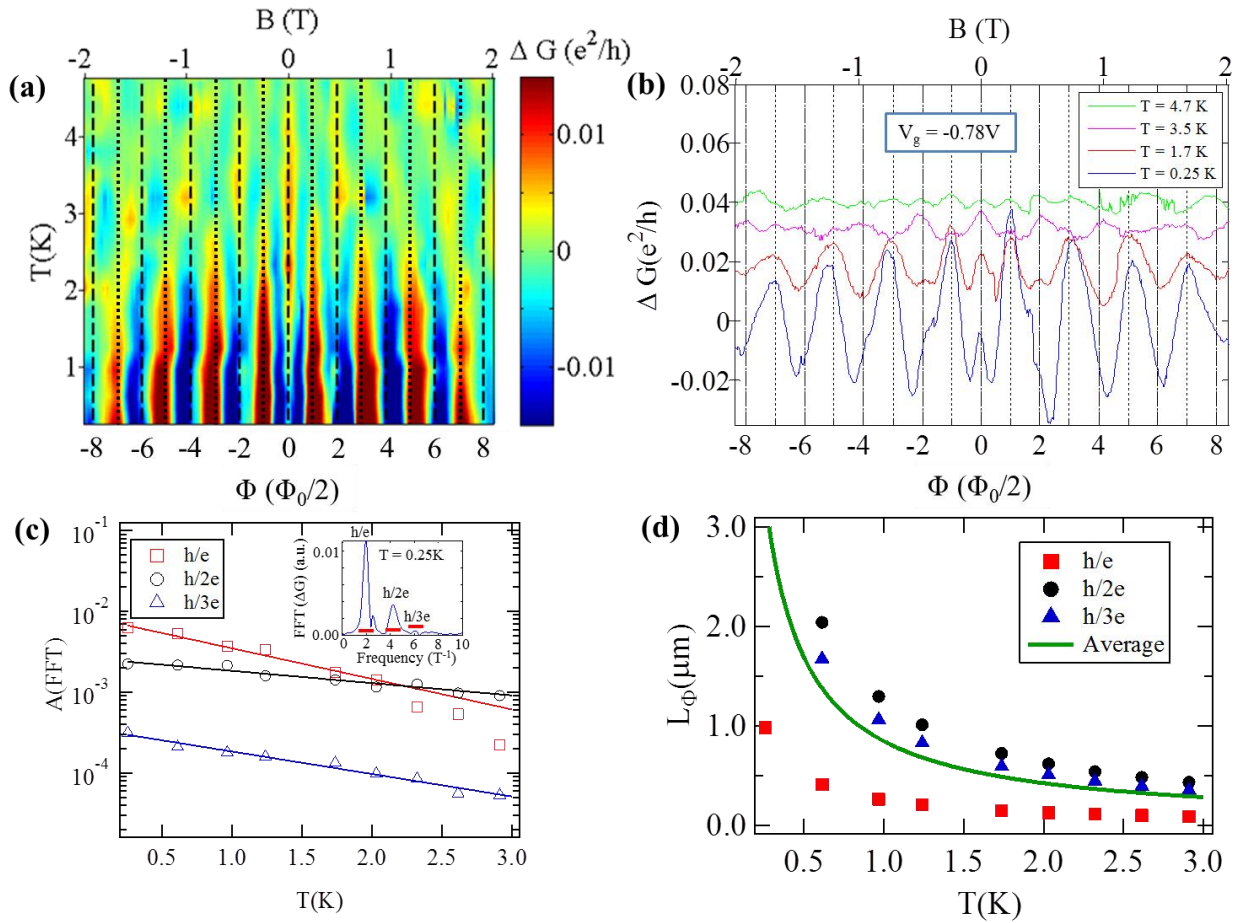

Figure 4. **Temperature dependence of the Aharonov-Bohm oscillations in TINRs, confirming quasi-ballistic transport.** **(a)** Color plot of $\Delta G$ (in units of $e^2/h$) vs. T and B (top axis, with corresponding magnetic flux $\Phi$ in bottom axis) for $V_g = -0.78V$. **(b)** $\Delta G$ vs. B (top axis, corresponding $\Phi$ in bottom axis) at different T's. For (a) and (b), vertical dashed lines represent $\Phi$ = even multiples of $\Phi_0/2$ (integer flux quanta), while dotted lines represent $\Phi$ = odd multiples of $\Phi_0/2$ (half-integer flux quanta). Curves are vertically offset for clarity. **(c)** Temperature-dependent amplitude of the FFT (A(FFT), in log scale) corresponding to the $h/e$, $h/2e$ and $h/3e$ peaks (inset) at $V_g = -0.78V$. Solid lines are exponential fits. Inset: Fast Fourier Transform (FFT) of $\Delta G$ vs. B at T = 0.25K. Each red horizontal line segment represents the interval over which the FFT is integrated to find the corresponding peak amplitude. **(d)** Temperature-dependence of the phase coherence length ($L_\Phi$) extracted from each FFT peak ($h/e$, $h/2e$, $h/3e$, plotted as the data in square, circle and triangles respectively, with the average between the 3 data sets shown as solid line). The exponential temperature dependence of A(FFT) and the observed three harmonics are consistent with quasi-ballistic transport[10,34].

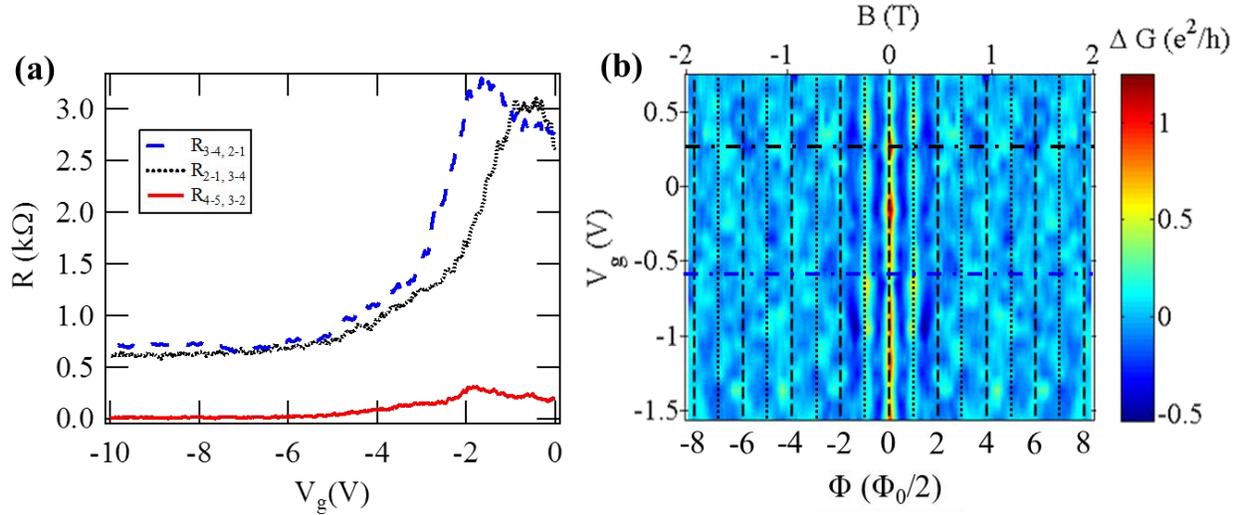

Figure 5. **Non-local transport in TINRs: Field effect and gate-tunable magnetoconductance oscillations.** **(a)** Field effect of the non-local resistance R measured with different sets of electrodes (i.e. $R_{2-1, 3-4}$ with current leads 2, 1 and voltage probes 3, 4, see device image in Fig. 2a). **(b)** Color plot of non-local $\Delta G$ (in units of $e^2/h$, measured with current leads 2-1 and voltage probes 3-4) vs. $V_g$ and B (top axis, corresponding magnetic flux $\Phi$ in bottom axis). Vertical dashed lines represent $\Phi$ = even multiples of $\Phi_0/2$ (integer flux quanta), while vertical dotted lines represent $\Phi$ = odd multiples of $\Phi_0/2$ (half integer flux quanta). Horizontal dashed dot lines are cuts at 2 representative $V_g$'s, shown in Fig. S5. All data are measured for T = 0.25K and I = 1nA.

**Supplementary Information**

The main text Fig. 1b shows the two principal momentum directions in a TINR, $k_{//}$ and $k_{\perp}$, which correspond to momentum parallel and perpendicular to the TINR length respectively. Because of the small NR circumference ($C$), $k_{\perp}$ is quantized to only take discrete values, with period $\Delta k = 2\pi/C$ (= 0.0015Å$^{-1}$ for our device). Furthermore, as pointed out in equation 1, in the presence of a magnetic flux ($\Phi$), $k_{\perp} = \Delta k \left( l + 0.5 - \dfrac{\Phi}{\Phi_0} \right)$, where l (angular momentum quantum number) takes only integer values.

The Fermi momentum ($k_F$) is the largest occupied momentum $k = \sqrt{k_{//}^2 + k_{\perp}^2}$, as shown in Fig. S1. The $k_F$ for the CNP is also depicted in Fig. S1 as $k_F^0$. Note whenever $k_F$ increments by $\Delta k$, the intercept of the Fermi surface (FS) on the $k_{\perp}$ axis would also increment by $\Delta k$, two new $k_{\perp}$ modes would become occupied (two new horizontal dashed lines intercept with the FS). Therefore we expect $\Delta G$ ($k_F$) to oscillate with period $\Delta k_F = \Delta k$. If we assign the first peak to oscillation index N = 0, the rest of peaks would occur at $k_F = k_F^0 + N\Delta k$ (N = 1,2,…). When the number of occupied $k_{\perp}$ *modes is large* (as in our TINRs, where even at CNP there are already >30 modes occupied as seen in Fig. 3c inset, note Fig. S1 here only depicts a few modes for clarity), $k_F$ can be approximately related with the 2D carrier density ($n_{2D}$) from the standard relation (for single spin Dirac fermion) as $n_{2D} = \dfrac{k_F^2}{4\pi}$.

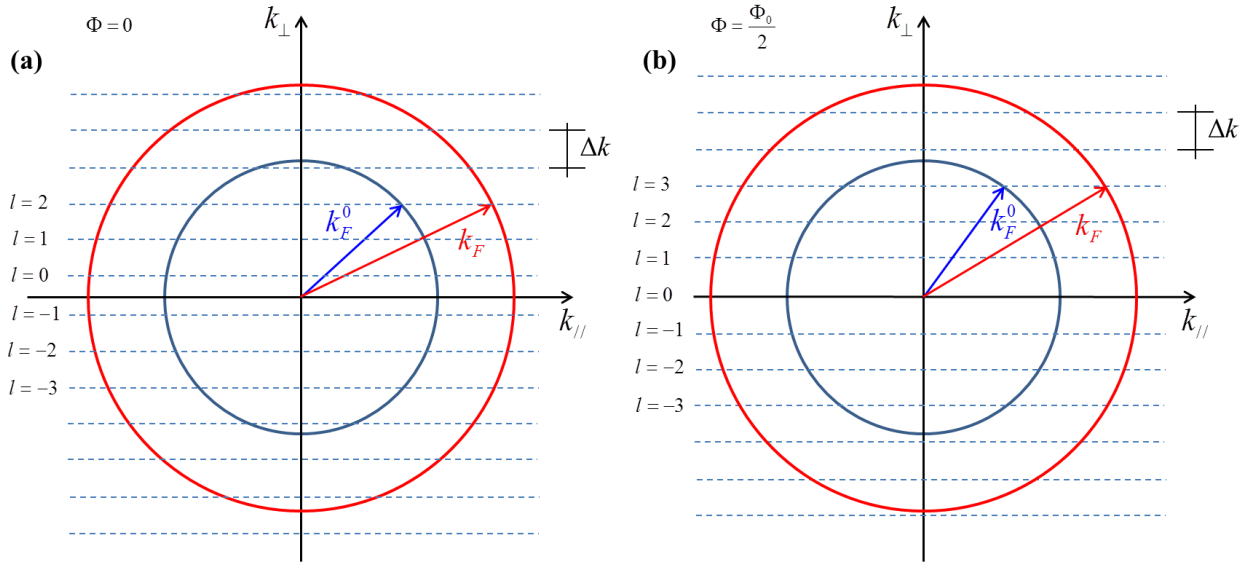

Figure S1. **Schematic of Fermi momentum in TINRs.** (a), (b) Fermi surface (circle), corresponding Fermi momentum ($k_F$, radius of circle), and formation of sub-bands (a series of E- $k_{//}$ dispersions (1D bands) indexed by discretely quantized $k_\perp$ as depicted in Fig. 1c, these surface (1D) sub-bands are also referred to as (circumferentially) quantized surface state modes) for the surface state in TINRs for two representative Φ's, Φ = 0 and Φ = $Φ_0/2$ respectively. The $k_{//}$ and $k_\perp$ are two surface momentum components parallel and perpendicular to the TINR axis (Fig. 1b). Horizontal dashed lines represent allowed discrete $k_\perp$, labeled with angular momentum quantum number (l), and spaced by $\Delta k = 2\pi / C$, where C is the TINR circumference. The red circle ($k_F$) is a representative Fermi surface (FS) at a finite n-type TSS carrier density, and the blue circle ($k_F^0$) is the FS at the charge neutrality point (CNP, for $k < k_F^0$ the TSS has energy lower than the top of BVB of $Bi_2Te_3$ thus not accessible in transport).

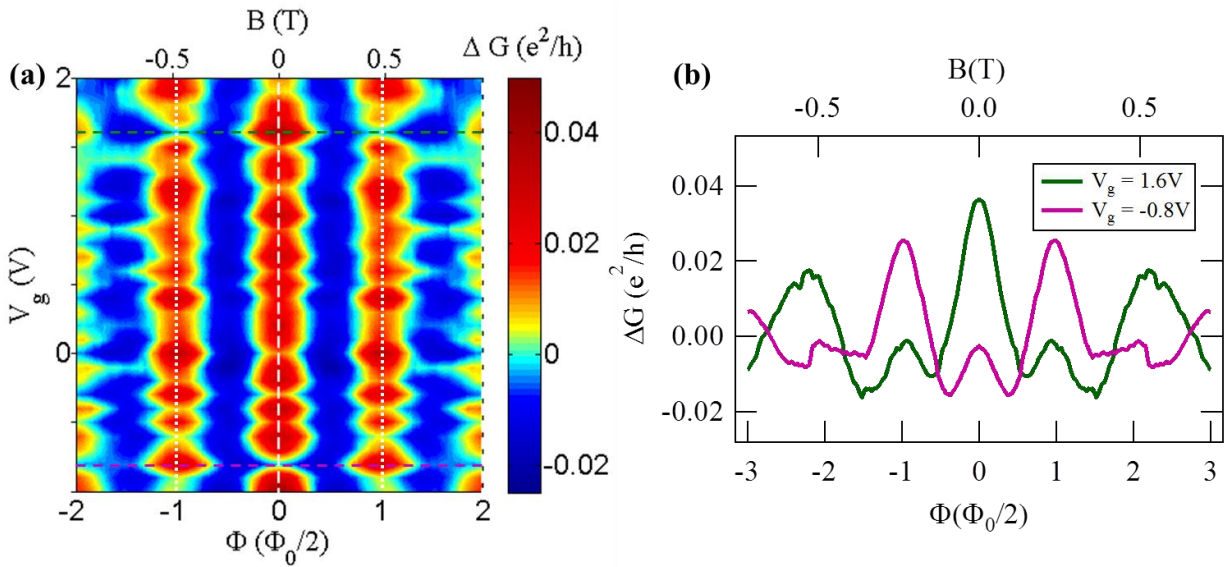

Figure S2. **Gate ($V_g$) and parallel magnetic field (B) dependent conductance (ΔG) and the 0/π-ABO alternations with $V_g$.** (a) A zoomed-in view of Figure 3a showing color plot of ΔG (in units of $e^2/h$) vs. $V_g$ (in 0.1V step, from -1V to 2V) and B (top axis, corresponding magnetic flux Φ in bottom axis, from –$Φ_0$ to $Φ_0$) measured at T = 0.25K. The vertical dotted lines represent half-integer flux quanta (Φ = ± $Φ_0/2$) and the vertical dashed line represent Φ = 0. Horizontal dashed lines are cuts at 2 representative $V_g$'s, shown in Fig. S2b. (b) Magnetoconductance (ΔG(B), with a smooth background subtracted), in units of $e^2/h$, vs. B-field (top axis, with corresponding magnetic flux Φ in units of half-flux-quantum ($Φ_0/2 = h/2e$) in bottom axis) at two representative $V_g$'s (horizontal cuts in Fig. 3a). The $V_g$ = -0.8V trace shows predominantly π-ABO while the $V_g$ = 1.6V trace shows predominantly 0-ABO. Magnetoconductance data have been symmetrized between opposite B field directions.

Figure S3a shows the temperature dependence of the $\Delta G_{6-1, 3-2}$ (B) vs. parallel B-field at $V_g$ = -1.2V. Figure S3b shows the horizontal cuts from Fig. S3a (color-coded dashed lines). The $\Delta G$ vs. parallel B at T = 0.3K (Fig. S3b) shows a predominant 0-ABO ($\Delta G$ peaks at even multiples of $\Phi_0/2$), however for T ~ 2.3K the π-ABO ($\Delta G$ peaks at odd multiples of $\Phi_0/2$) become more notable. For T ~ 4K, $\Delta G$ vs. B becomes predominantly 0-ABO again. Such a change in the phase of the ABO may be attributed to a temperature induced change of disorder potential in TINRs. Figures S3c and S3d present temperature dependence of the amplitude of FFT of ABO, and extraction of phase coherence length $L_\Phi$.

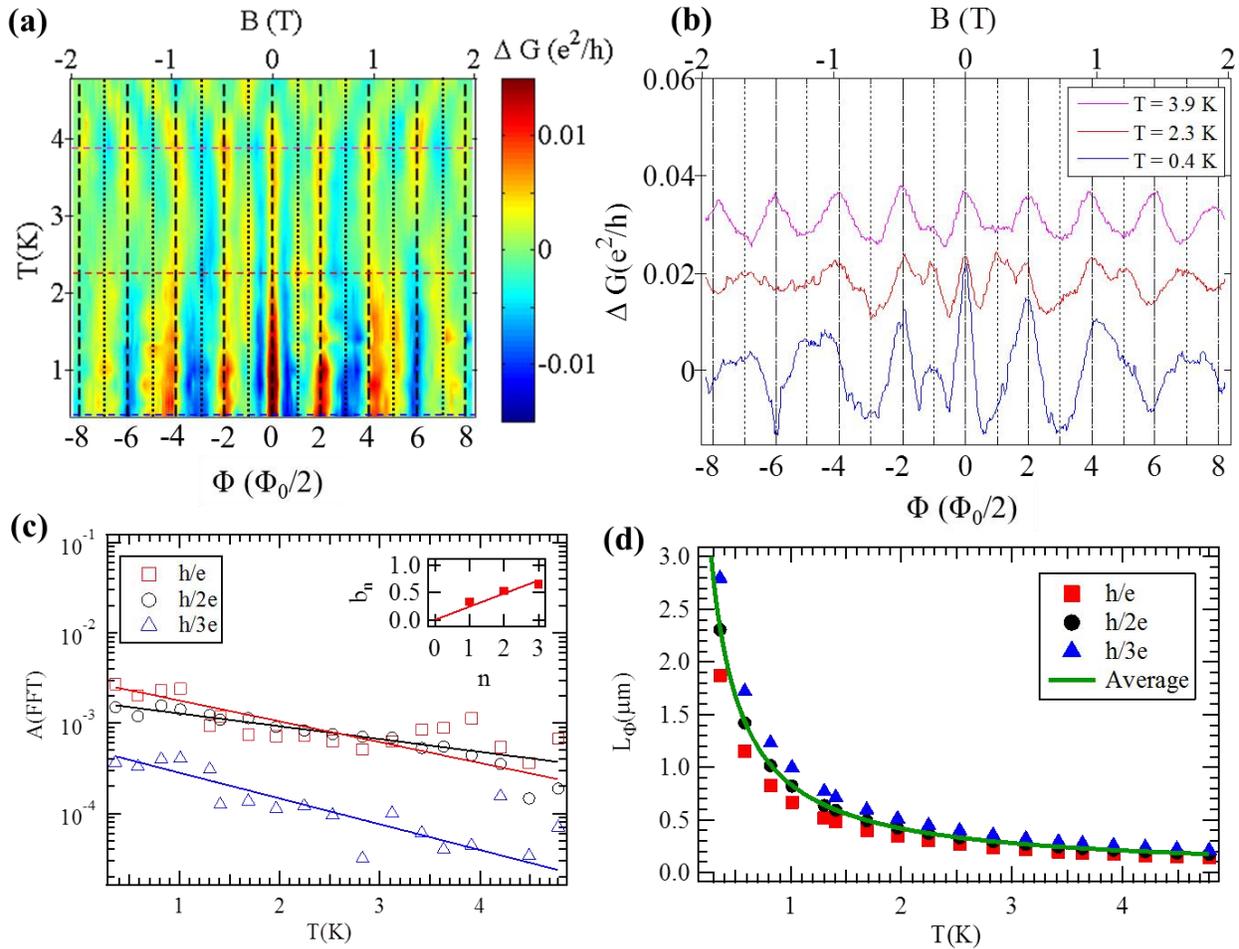

Fig. S3. **Temperature dependence of the Aharonov-Bohm oscillations amplitude.** (a) Color plot of $\Delta G$ vs. B (top axis, with corresponding $\Phi$ on bottom axis) and T at $V_g$ = -1.2V. Vertical dashed/dotted lines correspond to integer/half-integer flux quanta ($\Phi_0$) or even/odd multiples of $\Phi_0/2$. (b) $\Delta G$ (B) at representative T's (horizontal cuts, represented in (a) as color-coded dashed lines). Curves are vertically offset for clarity. (c) Temperature dependence of the amplitudes of the FFT peaks (A(FFT), in log scale) for the 3 corresponding periods (*h/e*, *h/2e* and *h/3e*). Solid lines are exponential fits. Inset: decay rates ($b_n$) vs. *n* (*n*th harmonic). (d) Temperature-dependent phase coherence length ($L_\Phi$) extracted from each FFT

peak ($h/e$, $h/2e$, $h/3e$, plotted as the data in square, circle and triangles respectively, with the average between the 3 data sets shown as solid line).

Figure S4 shows the 4-probes magneto-conductance $G_{6-1, 4-3}$ (B) vs. parallel B-field at $V_g$ = 1V. The channel length ($L_{ch}$) is $L_{ch}$ ~ 2.8 μm, around twice of $L_{ch}$ of $G_{6-1, 3-2}$ (shown in the main text). At large negative B field, we observe G peaks at $\Phi$ = even multiples of $\Phi_0/2$ ($h/2e$, 0-ABO) with period ~ 0.5T (consistent with the period measured for shorter channel length shown in main text), while at lower fields we observe coexisting/competing 0-ABO and π-ABO. The $L_\Phi$ extracted from the temperature dependence of the amplitude of ABO for a shorter channel length (Fig. 4 and Fig. S3) at T = 0.25K is ~ 3μm, similar to the channel length here, where it may be easier to lose coherence, explaining the less developed ABO in Figure S4 compared to Fig. 2.

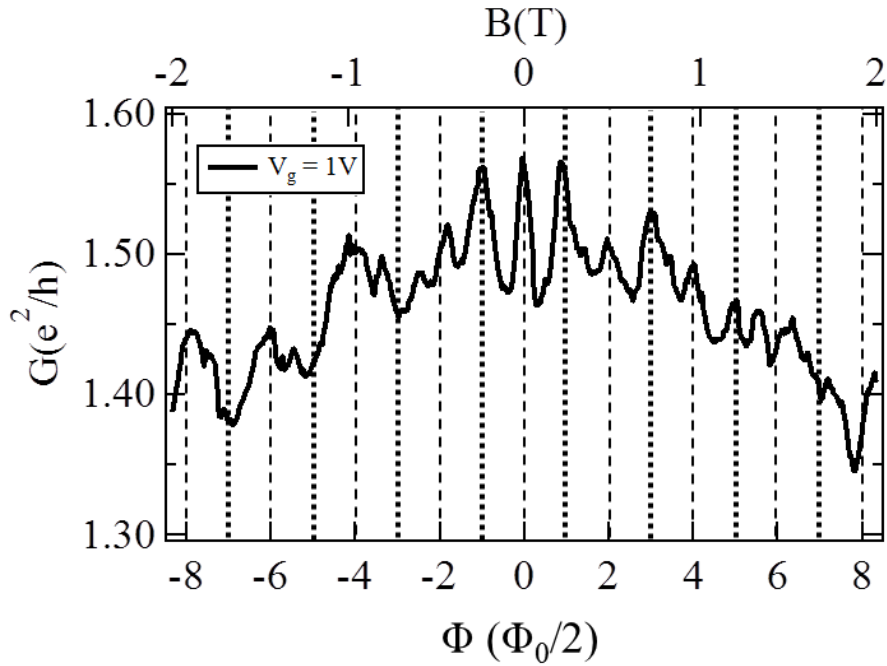

Figure S4. **Aharonov-Bohm oscillations in a longer TINR channel with length L = 2.8 μm.** Magneto conductance G ($G_{6-1, 4-3}$(B), in units of $e^2/h$) vs. B-field (parallel to NR axis) at $V_g$ = 1V at T = 0.25K. The scanning electron microscope (SEM) image of the device (with labeled electrodes) is depicted in the main text Figure 2a. The vertical dashed (dotted) lines represent integer units of flux quanta $\Phi_0$ or even multiples of $\Phi_0/2$ (half integer units of $\Phi_0$, or odd multiples of $\Phi_0/2$).

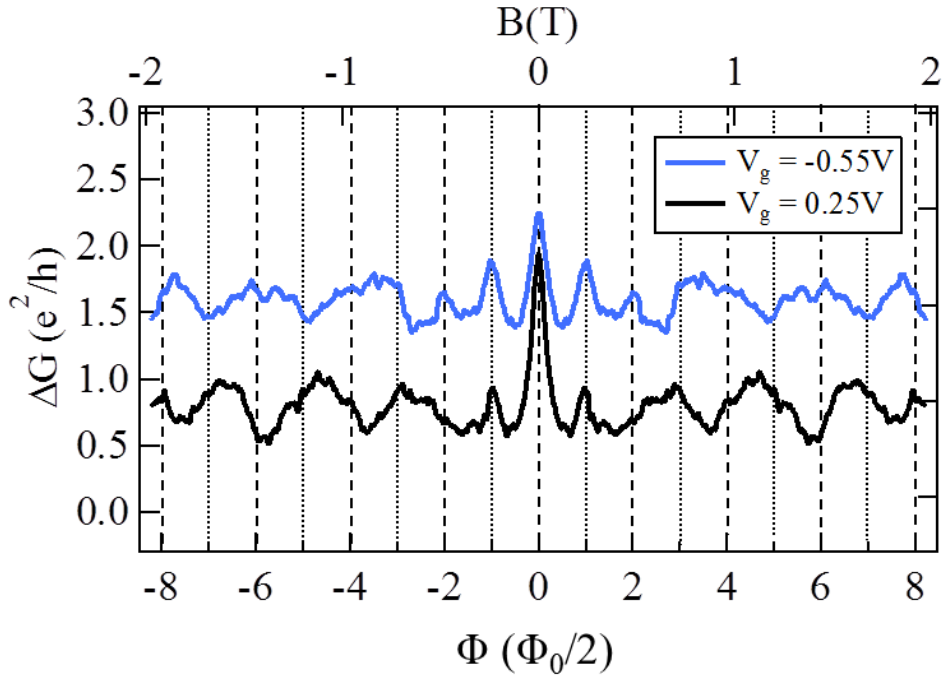

Fig. S5. **Gate-tunable, non-local magnetoconductance oscillations.** Nonlocal $\Delta G$ ($G_{2\text{-}1,\ 3\text{-}4}$ with polynomial background subtracted) vs. B (top axis, corresponding magnetic flux $\Phi$ in bottom axis) at different $V_g$'s (corresponding to the horizontal cuts in Fig. 5b) at T = 0.25 K. The vertical dashed (dotted) lines represent integer units of flux quanta $\Phi_0$ or even multiples of $\Phi_0/2$ (half integer units of $\Phi_0$, or odd multiples of $\Phi_0/2$). Curves are vertically offset for clarity.